\documentclass{article}
\usepackage{graphicx} 

\title{Natural Intelligence: the information processing power of life}
\author{Seth Lloyd, Michele Reilly, MIT}
\date{May 2025}

\begin{document}

{\large
\maketitle
\begin{abstract}

Merely by existing, all physical systems contain information, and physical dynamics transforms and processes that information.
This note investigates the information processing power of living systems. 
Living systems harvest free energy from the sun, from geothermal sources, and from each other.  They then use that free energy to drive the complex set of chemical interactions that underlie life.  All molecules --  be they simple molecules such as water, or complex molecules such as DNA -- register information via their chemical composition.  When these molecules undergo chemical reactions, that information is transformed and processed.   These chemical transformations can be thought of as elementary logical operations: such bio-ops include the absorption of a photon in a chromophore during photosynthesis, the formation or breaking of covalent, hydrogen, and van der Waals bonds in the process of metabolism and reproduction, or the release of a neurotransmitter molecule when a synapse fires in the brain.   This paper estimates the total number of bio-ops that have been, and are being performed, by life on earth.  We find that the current number of bio-ops performed by all life on earth is around $10^{33}-10^{35}$ bio-ops per second.  The cells in an individual human being perform around $10^{20}-10^{22}$ bio-ops per second, comparable to the information processing power of all the computers, cell phones, and server farms on earth.   Depending on how one defines a neural operation, at most a few percent of human bio-ops take place in the firing of neurons and synapses in the brain. Over the course of life on earth, about $10^{50}-10^{52}$ bio-ops have taken place.
\end{abstract}
\section{Introduction}

The computational capacity of physical systems is governed by the fundamental physics of computation [1-2].    All physical systems register information -- the information that specifies the microscopic state of the molecules, atoms, and elementary particles that make up the system.   When those molecules, atoms, and elementary particles interact, that information is transformed and processed.   The universe computes.   

This physical information processing obeys simple laws [1-2].  The maximum amount of information that can be stored by a physical system is equal to its maximum entropy.  To flip a bit in time $t$ requires energy $\pi\hbar/2E$, where $E$ is the energy of the bit above its ground state.  Existing quantum computers saturate these physical bounds. The physical laws of computation can be used to bound the information processing power of various systems, including human-made computers, black holes, and the universe itself, which can have performed no more than  $10^{120}$ operations on $10^{120}$ bits [2].

A recent paper [3] uses these fundamental physical bounds to estimate the computational capacity of all life on earth.  The computational power of life is clearly a sensible thing to calculate, not least because the calculation allows us to compare the information processing power of living systems with that of human artifacts such as electronic computers.   As will be seen, the amount of information processing performed by living systems in the process of harvesting energy, metabolism, and reproduction is vastly greater than that performed by computers, which are much less efficient than living systems in their energy use.

In trying to verify the results presented in [3], we came up with rather different numbers for how much information processing has been performed -- and is being performed -- by life. We invite the reader to compare the methods used here with the methods used in [3].  The current paper presents our estimates obtained from applying physical limits to computation to the wide variety of information processing tasks performed by living systems.

\section{Computational capacity of life}

As noted, all molecules register information in their chemical composition and shape.  For the purposes of this paper, we will define a bio-op to be an elementary molecular change that has a biologically active function. 
Bio-ops then include the breaking or formation of a chemical bond -- covalent, hydrogen or van der Waals -- in the course of a biochemical process, the absorption of a photon in the antenna of the photosynthetic apparatus, the synthesis of a molecule of ATP, the absorption of a neurotransmitter molecule in a synapse, or the copying of DNA.  

A simple way to estimate the number of bio-ops performed by a living system is to look at the total number of chemical bonds formed and broken in biological reactions on earth.     
This energy comes primarily from the sun, which shines about  $1.7 \times 10^{17}$ watts of power on the whole earth [4].   (Geothermal energy also contributes free energy to the living process, but the total geothermal heat flux from the interior of the earth is considerably less than solar energy -- on the order of $5 \times 10^{13}$ watts [5].)  Losses from reflection and from absorption in the atmosphere, the ground, and the ocean, reduce the amount of sunlight available to living systems to approximately $10^{16}-10^{17}$ watts.   The fraction of solar energy captured as chemical energy via photosynthesis is $1-3\%$ [6]:  all in all, around $10^{14}-10^{15}$ watts of the solar power falling on earth is captured as the chemical energy that fuels living cells. 

We define a bio-op as the formation or breaking of a chemical bond in the metabolic and reproductive process of a living system, including photosynthesis.
Covalent bonds have energy $1-3$ electron volts, where one electron volt $ = 1.6 \times 10^{-19}$ joules.    Hydrogen bonds have energy around an order of magnitude smaller, and the energies of van der Waals interactions are two to ten times smaller still.  Since one electron volt $ = 1.6 \times 10^{-19}$ joules, we obtain $10^{33} -- 10^{35}$ bio-ops performed per second in all living matter.

Life on earth has been around for $\approx 4 \times 10^9$ years, and there are approximately $\pi \times 10^7$ seconds per year.  We can then estimate that circa $10^{50}-10^{52}$ bio-ops have been performed over the history of life on earth.  This number is
ten orders of magnitude larger than the lower bound suggested in [3], which is based on counting super-radiant photons in cytoskeletal fibers, and eight to ten orders of magnitude smaller than the upper bound suggested in [3], which is based on the assumption that the entire $E=mc^2$ mass energy of biological systems is devoted to computation.   By contrast, we count only the energy devoted to the breaking and formation of bonds in bio- and photo-chemical reactions.   

The metabolic rate of human beings is on the order of 100 watts (ca. 100 kcal/hour), of which perhaps 1-10\% -- given the inefficiencies of the metabolic process -- goes directly into performing bio-ops.   Since a bio-op uses between $O(0.1 -1)$ eV, we find that the number of bio-ops being performed by the cellular processes that underlie any living human being's existence is around
$10^{20} - 10^{22}$ bio-ops per second.    There are around ten to one hundred billion digital processors in the world (including computers, smart phones, servers, etc.), each performing around ten to one hundred billion logical operations per second,
leading to an estimate of $10^{20} - 10^{22}$ ops per second [7].  The number of bio-ops going on in your body is on the same order of magnitude as the number of logical operations being performed by all the human-made electronic computers in the world.

It is salutary to compare the amount of information processing going on in all of our cells, with the amount of information processing in the brain.  There are different ways of estimating the amount of information processing in the brain: in the spirit of defining a bio-op as the formation or breaking of a chemical bond, we define a neural bio-op to be the release or binding of a neurotransmitter molecule to a receptor during the firing of a synapse.  The brain has around $10^{11}$ neurons, with an average firing rate of around 10 times per second.  Each neuron has on the order of $10^{3}$ synapses, each of which releases on the order of $10^4$ neurotransmitter molecules when the neuron fires.   The synaptic binding energy is on the order of half an electron volt per neurotransmitter.  If we measure the total number of bio-ops in the neural dynamics in terms of the number of synaptic binding events per second, we obtain around $10^{19}$ bio-ops per second in the operation of the brain [8].   (If we count only the information contained in neural spike trains, we obtain a few tera-ops per second.)  That is, the brain is performing at most a few percent of the bio-ops performed by the body as a whole. 

\section{Natural intelligence vs. artificial intelligence}

Most of the information processing taking place in our bodies is performed by cells taking in information and free energy and using that energy to process information by performing the complex series of chemical reactions that allow those cells to live and to reproduce.   This natural intelligence has been honed by billions of years of evolution to evolve highly sophisticated computational mechanisms.   Because of the evolutionary advantage accrued by using less energy to perform bio-ops, computation at the cellular level is highly efficient compared both with neural information processing and with human-made digital computers.  
To compare, the data centers of the United States consume ten times more energy (ca. half a terawatt [9]) than the inhabitants of the USA consume in food (ca. 100 watts/person).    And yet {\it one individual human being} possesses more information processing power, measured in bio-ops, than all the data centers in the world.

Moreover, the data by which physical intelligence was trained consists of the experience of billions of years of evolution: as shown above, ca. $10^{50}$ bio-ops of living computation went into honing the biochemical information processing mechanisms that underlie life.  

By contrast, current large language models have been trained on datasets containing on the order of $10^{14} - 10^{15}$ bits of information, consisting of  the `common crawl' (web pages: ca. 50\% of the training data), news sources/forums (20\%), books and literature (15\%), scientific papers (10\%), and Wikipedia (5\%) [10].     Impressive as recent advances in AI are, the raw numbers suggest that it is perhaps too soon to expect artificial intelligence to exhibit the efficiency and sophistication of natural intelligence as manifested in, e.g., the operation of the central nervous system together with its sensory apparatus, or the cellular DNA repair mechanism.

\section{The language of nature and the physical 
Church-Turing hypothesis}

The primary difference between natural intelligence and artificial intelligence is that the foundation of natural intelligence is the language of nature, whereas the foundation of artificial intelligence is human language.   The laws of nature -- physics, chemistry, biology -- program natural intelligences to generate and to recognize complex patterns in the universe.   Information processing at the cosmic scale, mediated by the laws of gravitation and thermodynamics, creates planets, stars, galaxies, and clusters of galaxies.    Due to a quirk of gravity, the increase of entropy dictated by second law of thermodynamics, instead of leading to homogeneous systems at the same temperature, generates structures at all cosmic scales, and concentrates, rather than disperses free energy.   Hydrogen atoms cluster together to form stars, releasing their energy via nuclear fusion.    The flux of free energy from hot stars to cold space via the earth provides the computational fuel for life.

Just what structures does the language of nature generate?   The physical Church-Turing hypothesis [11] conjectures that the structures generated by the language of nature are those that can be generated by a universal quantum computer.  This remarkable conjecture  -- consistent with our current understanding of the laws of physics and computation -- links the creations of natural language to those of human language. 

Human language is one of the most profound discoveries of natural intelligence.  Human language possesses a universal quality not apparently found in the language of other animals. The essence of this universality lies in what is 
called `recursion.' Recursive language possesses the capacity both for self-reference, and for open-ended construction of ever more complex structures.   All human languages possess this universal recursive ability, and crucially, we have gifted this ability to artificial intelligences via computer languages.  Indeed, the universal properties of recursive language were discovered and made precise in 1936 by Alan Turing's invention of the concept of the digital computer [12], and by Alonzo Church's invention of the first universal computer language, the Lambda Calculus [13]. 

The Church-Turing hypothesis conjectures that any mathematical function that can be computed, can be  computed by a universal digital computer.  In 1985, David Deutsch formulated the physical Church-Turing hypothesis [11], which states that any structure found in nature can be efficiently simulated on a universal {\it quantum} computer.  
The physical Church-Turing hypothesis is consistent with the laws of physics and of computation in the sense that there are no known examples of physical or mathematical structures that can not in principle be generated and efficiently simulated by a quantum computer.    Indeed, given our current understanding of the laws of physics, any quantum system with bounded volume and energy can in principle be simulated accurately and efficiently by a quantum computer [1-2].

The language of nature underlies human and computer language.  Natural law enables recursive language and computation: my brain is a complex of electrochemical reactions, and I am writing this on a digital computer whose dynamics is governed by the laws of physics.   Natural language and its offspring, recursive thought, are open-ended: as Turing showed [12], we cannot know in advance what a computer will output when it follows the instructions of any given program, or indeed whether it will output anything at all.  Natural, human, and artificial intelligence are powerful, and not intrinsically benign.   The computationally open-ended nature of Nature makes intelligence also inherently unpredictable.

\section{Conclusion}

This paper has used the physics of computation to estimate the information processing power of biological systems.   A logical operation in an electronic digital computer is performed when a transistor switches from one logical/physical state to another in the ongoing process of computation.   We have defined a bio-op to be a microscopic bio-physical transformation such as the formation or breaking of a chemical bond in the ongoing process of biological computation that underlies life.   We found that biological systems process large amounts of information -- much larger than that processed by all the digital computers on earth.

Further comparing natural vs. artificial intelligence, we noted that the training set for natural intelligence -- the information processed during the whole history of natural selection on earth -- is vastly larger than the training sets used by artificial intelligence.   The exhaustion of available digital training sets already poses challenges for large language models.   These challenges will be exacerbated as we try to extend the power of artificial intelligence to systems such as robots and self-driving cars that must navigate the physical world. 

Human language and recursive thought are remarkable manifestations of natural intelligence.  We have in turn gifted the powerful capacities of recursive reasoning to digital computers.  The physical Church-Turing hypothesis suggests that this computational ability (extended to the quantum realm) gives us and our computers the capacity -- in principal -- to understand and to reconstruct the behavior of all physical systems.  Artificial intelligence is a manifestation of natural intelligence: it is the latest link to be forged in the `great chain of computation' that binds together the universe. 
\vfill

\noindent{\it Acknowledgements:} SL would like to thank Philip Kurian for productive discussions.

\vfil\eject

\noindent{\it References:}

\bigskip\noindent[1]
Lloyd S 2000 Ultimate Physical Limits to Computation {\it Nature} {\bf 406} 1047-1054.

\bigskip\noindent[2]
Lloyd S 2002 Computational Capacity of the Universe
{\it Physical Review Letters} {\bf 88} 237901.

\bigskip\noindent[3] Kurian P 2025 Computational capacity of life in relation to the universe 
{\it Science Advances} {\bf 11} eadt4623. 

\bigskip\noindent [4] Trenberth KE, Fasullo JT, Kiehl J 2009
Earth's Global Energy Budget.
{\it Bulletin of the American Meteorological Society,} {\bf 90}(3), 311–324.
https://doi.org/10.1175/2008BAMS2634.1

\bigskip\noindent
[5] Davies JH and Davies DR 2010 
Earth's surface heat flux
{\it Solid Earth} {\bf 1(1)} 5–24. 
https://doi.org/10.5194/se-1-5-2010.

\bigskip\noindent
[6] Blankenship RE 2021 Molecular Mechanisms of Photosynthesis, Wiley.

\bigskip\noindent
[7] This simple estimate of the current global information processing power of computers -- ca. $10^{21}$ ops per second, or one zetaflop -- yields similar order of magnitude results as other methods of estimating global computing power (e.g., total global computing power consumption 
$\approx 10^{11}$ watts times $10^{10}$ ops/watt $\approx 10^{21}$ ops/second).   For a detailed assessment of current and projected computer power, see {\it The International Roadmap for Systems and Devices IEEE 2022}.

\bigskip\noindent
[8] Principles of Neural Science (6th Edition, 2021)
ER Kandel, JD Koester, SH Mack, SA Siegelbaum, eds.
McGraw-Hill Education
ISBN: 978-1259642234.
If one takes a more macroscopic version of bio-ops in the brain -- identifying a bio-op with a neuron or a synapse firing, rather than a neurotransmitter being released or binding to a synaptic receptor -- one obtains lower numbers for the brain's information processing power: from $10^{15}$ to $10^{18}$ operations per second. 

\bigskip\noindent
[9] 
Shehabi A, Smith SJ, Hubbard A, Newkirk A, Lei N, Siddik, MAB, Holecek B, Koomey
J, Masanet E, Sartor D  2024 United States Data Center Energy Usage Report.
Lawrence Berkeley National Laboratory, Berkeley, California. LBNL-2001637

\bigskip\noindent
[10] Open AI {\it et al.} GPT-4 Technical Report 2023 arXiv:2303.08774.

\bigskip\noindent
[11] Deutsch D 1985 Quantum theory, the Church–Turing principle and the universal quantum computer {\it Proc. R. Soc. Lond. A} {\bf 400} 97–117.
http://doi.org/10.1098/rspa.1985.0070

\bigskip\noindent
[12] Turing AM 1936 On computable numbers, with an application to the Entscheidungsproblem {\it Proceedings of the London Mathematical Society} {\bf 42}(1) 230-265.

\bigskip\noindent
[13] 
Church A 1936
An Unsolvable Problem of Elementary Number Theory
{\it American Journal of Mathematics} {\bf 58} 345–363.

}
\end{document}